\documentclass[aip,reprint]{revtex4-1}

\usepackage{graphicx}
\usepackage{dcolumn}
\usepackage{bm}

\usepackage[utf8]{inputenc}
\usepackage[T1]{fontenc}
\usepackage{mathptmx}
\usepackage{etoolbox}

\usepackage{color}

\usepackage{amssymb}
\usepackage{siunitx}

\draft 

\begin{document}


\title{Rheology of Polydisperse non-Spherical Graphite Particles Suspended in Mineral Oil} 



\author{Th. Larsen}
\author{A. L. Søbye}
\affiliation{Department of Materials and Production, Aalborg University, 9000 Aalborg, Denmark}
\affiliation{Advent Technologies A/S, 9000 Aalborg, Denmark}

\author{J. R. Royer}
\author{W. C. K. Poon}
\affiliation{School of Physics and Astronomy, University of Edinburgh, Edinburgh EH9 3FD, United Kingdom}

\author{T. Larsen}
\author{S. J. Andreasen}
\affiliation{Advent Technologies A/S, 9000 Aalborg, Denmark}

\author{A. D. Drozdov}
\author{J. D. C. Christiansen}
\email[Author to whom correspondence should be addressed; electronic mail:  ]{jc@mp.aau.dk}
\affiliation{Department of Materials and Production, Aalborg University, 9000 Aalborg, Denmark}

\date{\today}

\begin{abstract}
We study the role of filler concentration and microphysics on the rheology of polydisperse flake-graphite particles suspended in Newtonian mineral oil. Under steady shear, our samples exhibit shear thinning and yielding behaviour is observed for volume fractions $\phi > 0.18$. Time-temperature superposition was observed using an Arrhenius-type horizontal shift factor, giving a flow activation energy that is dependent on the graphite volume fraction, suggesting concentration-dependent contributions to relaxation processes in the suspensions. The flow curves are fitted by a constraint-based model, indicating that the flow behaviour is controlled by frictional and adhesive contacts, with the model suggesting that the adhesive stress is temperature dependent. 
\end{abstract}


\maketitle 


\section{Introduction}

Carbonic polymer composites are widely applicable, for instance as electrode materials in fuel cell bipolar plates (BPPs), as porous electrodes in Li-ion batteries, or in aerospace composites, and hence, their properties have been intensively studied  \cite{Zhang2005,Naz2016,Yao2017,Sengupta2011,Kakati2009,Mendes2007,Lim2015}. In BPP applications, replacing traditional materials such as graphite with highly-filled carbonic polymer composites can give higher electrical and thermal conductivities at relatively low costs \cite{Lee2009}. The high filler loadings required to obtain satisfactory conductivities may, however, dramatically affect the composite rheology \cite{King2006,King2008}, hereby complicating manufacturing. Yet, characterising the rheological properties of carbonic polymer composites may yield valuable insight into the microstructure, filler dispersion and filler interactions since this information correlates with electrical properties \cite{daSilva2021}. Thus, rational composite design requires understanding how the filler particles, the matrix phase and their coupling affect the system's rheology. Additionally, the importance of understanding the coupling between phases extends to e.g. semi-solid flow cells where the flowing electrodes consist of suspensions of active (LiFePO$_4$) and conductive (carbon black) materials. Tuning particle interactions in such biphasic mixtures allows control of their rheology and hence reduction of particle sedimentation and phase segregation, as well as enabling efficient charge transport \cite{Wei2015}.

The matrix in graphite-polymer composites is typically a thermoplastic such as poly(methyl-methacrylate), polyethylene, polyvinyl chloride, etc., and their derivatives. The rheology of their melts at processing temperatures is by now well understood~\cite{doi1988,McLeish2002}. This provides a good starting point for the predictive formulation of the matrix component of graphite-polymer composites. 

The filler, graphite, can be exfoliated into colloidal flakes~\cite{Ma2015,Zheng2012}. However, most commercial graphite powder additives fall in the non-Brownian regime, with typical particle size $\gtrsim \SI{10}{\micro\metre}$. Compared to polymer melts, the rheology of non-Brownian suspensions is poorly understood. Such systems are history-dependent because no thermal motion randomises particle positions. Moreover, it has recently become clear that particle contacts control rheology~\cite{Morris2020}, and contact forces are far more sensitive to surface details, e.g., hydrogen bonding~\cite{James2018}, than non-contact colloidal interactions. Many uncertainties therefore remain, with experimental validation of simulations~\cite{Mari2014} and theory~\cite{Wyart2014} largely limited to model non-Brownian suspensions of nearly-monodispersed repulsive spheres~\cite{Guy2015, Royer2016, Hermes2016}, although results for the role of adhesive contact interactions are now beginning to emerge~\cite{Guy2018,Richards2020,Richards2021}. Not surprisingly, then, highly non-intuitive discoveries are still being made, such as bistability in non-Brownian particles dispersed in a colloidal gel matrix~\cite{Jiang2022}. 

Current knowledge therefore does not yet constitute a firm basis for the predictive formulation of graphite-polymer composite fillers. In this work, we contribute towards this goal by studying a model composite in which the non-Newtonian polymer melt matrix at processing temperatures is replaced by a simpler Newtonian fluid. We then probe whether the recently-emergent framework of `contact-force rheology'~\cite{Morris2020,Guy2018} may be applicable to one popular form of graphitic particles. 

Specifically, we explore the rheology of polydisperse graphite particles dispersed in mineral oil using rotational rheometry, varying the solid volume fraction, $\phi$, and temperature, $T$. A weak flow-history dependence may be attributable to filler orientation. We re-scale flow curves from different $T$, and fit these with the purely empirical Herschel-Bulkley model and a recent physics-based model. The latter approach allows us to infer mechanistic insights concerning adhesive and frictional inter-particle contacts. As background, we first review briefly the new framework of suspension rheology involving inter-particle contacts.

\section{Contact forces in suspension rheology}

Developing constitutive models for non-Brownian, or granular, suspensions is a longstanding challenge \cite{Ancey1999}. Recent advances have identified an essential piece of missing physics. Simulations~\cite{Mari2014} and theory~\cite{Wyart2014} suggest that at high enough applied stress, particles are pressed into frictional contact. This effect is included in a simple model~\cite{Wyart2014} that starts from the well-known phenomenological equation~\cite{Mueller2010,Maron1956,Krieger1959}
\begin{equation}
    \eta_r = \left(1 - \frac{\phi}{\phi_J} \right)^{-l},
    \label{eq:results:Krieger-Dougherty_model}
\end{equation}
with $\eta_r= \eta/\eta_0$ the suspension viscosity, $\eta$, relative to that of the suspending solvent, $\eta_0$, and the exponent\footnote{Note that in the Krieger-Dougherty equation the exponent was chosen as $l = \left[\eta\right]\phi_J$, with $\left[\eta\right]$ the intrinsic viscosity, to give the correct first-order term for a Taylor expansion in the dilute limit. While this happens to give a reasonable numerical value for $l$ for spherical particles, there is no fundamental physical reason for the intrinsic viscosity, describing single-particle hydrodynamic contributions, to control the divergence near jamming.}  $l \approx 2$. This so-called Krieger-Dougherty equation captures the well-known observation that the viscosity of any suspension will diverge when its volume fraction, $\phi$, reaches some `jamming' point $\phi_J$. In the original formulation, $\phi_J$ is a constant for any particular suspension. Subsequently, it has been suggested that $\phi_J$ is shear-rate dependent~\cite{Wildemuth1984, Stickel2005}. The recent breakthrough comes from the realisation that $\phi_J$ is in fact dependent on the applied stress, $\sigma$, for which Wyart and Cates (WC) propose the form
\begin{equation}
  \phi_J(\sigma) = \phi_{\rm rlp}f(\sigma) +\phi_{\rm rcp}\left[1-f(\sigma)\right].
    \label{eq:introduction:WC-phiJ}
\end{equation}
This is a linear interpolation between random close packing, $\phi_{\rm rcp}$ in the low-stress limit and random loose packing, $\phi_{\rm rlp}<\phi_{\rm rcp}$, in the high-stress limit. The crucial physics idea is that as the applied stress increases, an increasing fraction $f$ of particles are pressed into frictional contact. The no-sliding constraint imposed by static friction then dictates a looser form of packing, so that as $f$ increases from 0 to 1, the jamming point $\phi_J$ decreases from random close packing to random loose packing, with the latter being dependent on the coefficient of static friction, $\mu$, between the particles. For non-frictional hard spheres, $\phi_{\rm rcp} \approx 0.64$, and  $\phi_{\rm rlp}\approx 0.55$ in the high friction limit ($\mu\gtrsim~1$)~\cite{Silbert2010}. 
WC propose that the transition between these two packing limits occurs at some characteristic `onset stress' $\sigma^*$ for frictional contact. Any form of $f(\sigma)$ that increases with $\sigma$ predicts shear thickening when Equation \ref{eq:introduction:WC-phiJ} is substituted into Equation \ref{eq:results:Krieger-Dougherty_model}: $\eta_r$ increases as $\sigma$ increases because $\phi_J$ decreases with $\sigma$. A sigmoidal $f(\sigma)$ captures the measured rheology of a variety of shear thickening systems with only a handful of fitting parameters \cite{Guy2015, Royer2016, Hermes2016}.

This framework can be extended to include adhesive contacts that constrain rolling,  which are released under increasing stress~\cite{Guy2018}. Such rolling constraints, which could arise between attractive particles due to contact pinning or touching facets, introduce a new critical packing fraction, `adhesive loose packing', $\phi_{\rm alp}<\phi_{\rm rlp}$. This limit is not precisely known even for monodisperse spheres, although simulations~\cite{Liu2017} suggest that it could be as low as  $\phi_{\rm alp}\approx 0.14$. Analogous to $f(\sigma)$, a function $a(\sigma)$ describes the fraction of adhesive constraints, with $a(\sigma\to0)= 1$ and $a(\sigma\to\infty)= 0$. The extended WC model including such adhesive contacts predicts a `zoo' of flow curves involving combinations of shear thinning and shear thickening~\cite{Guy2018}.

In the limit where contacts are always frictional\footnote{In such a system, shear thickening is not observed; in a loose sense, the ever presence of friction means that flow is always frictional, so that the system is `always shear thickened'.}, i.e. $\sigma^*\to 0$ and $f = 1$ at all $\sigma$, we write
\begin{equation}
  \phi_J(\sigma) = \phi_{\rm alp}a(\sigma) +\phi_{\rm rlp}\left[1-a(\sigma)\right].
    \label{eq:introduction:ADH-phiJ}
\end{equation}
An inverse sigmoidal form 
\begin{equation}
    a(\sigma) = 1 - \exp{\left[-\left(\sigma_a/\sigma \right)^{\beta} \right]}
    \label{eq:introduction:a_sig}
\end{equation}
is able to capture the rheology of a variety of systems, including PMMA spheres in oil \cite{Guy2018}, cornstarch in oil \cite{Richards2020}, calcite in glycerol-water \cite{Richards2021} and molten chocolate (= sugar grains in oil)~\cite{Blanc2018}. In Equation \ref{eq:introduction:a_sig}, $\sigma_a$ gives a characteristic stress scale for peeling adhesive contacts apart to initiate rolling, while $\beta$ describes the rate $a(\sigma)$ decreases from 1 to 0. Importantly, a finite yield stress, $\sigma_y > 0$, is predicted at $\phi>\phi_{\rm alp}$. At and below this stress, the viscosity diverges in the flow curve. Solving $\eta^{-1}(\sigma) = 0$ using the above ansatz for $a(\sigma)$ gives 
\begin{equation}
    \sigma_y(\phi) = \sigma_a \left[\ln \left( \frac{\phi_{\rm rlp} - \phi_{\rm alp} }{\phi-\phi_{\rm alp}} \right) \right]^{-1/\beta},   \label{eq:introduction:sigY}
\end{equation}
which describes $\sigma_y(\phi)$ data in calcite suspensions~\cite{Richards2021}. Note that the yield stress involves the cooperative action of adhesion {\it and} friction~\cite{Richards2020}, with adhesive contacts stabilising frictional force chains. Moreover, Equation \ref{eq:introduction:sigY} predicts $\sigma_y \to \infty$ at $\phi_{\rm rlp}$, just as in purely frictional systems: irrespective of adhesion, friction prevents flow above random loose packing.  
   
In this model, many effects enter through the values of $\phi_{\rm alp}$ and $\phi_{\rm rlp}$. In particular, their values depend on  particle shape and polydispersity: irregular particles jam at lower volume fractions compared to spheres \cite{Torquato2010} while polydispersity enables denser packing \cite{Farr2009}. Moreover, this simple model will not account for effects such as shear-induced alignment of anisotropic particles. Nevertheless, we will find fitting this model to our graphite-in-oil data  gives significant insight into the physical factors controlling the rheology of this system.

\section{Materials and Methods}

\begin{figure}
    \centering
    \includegraphics{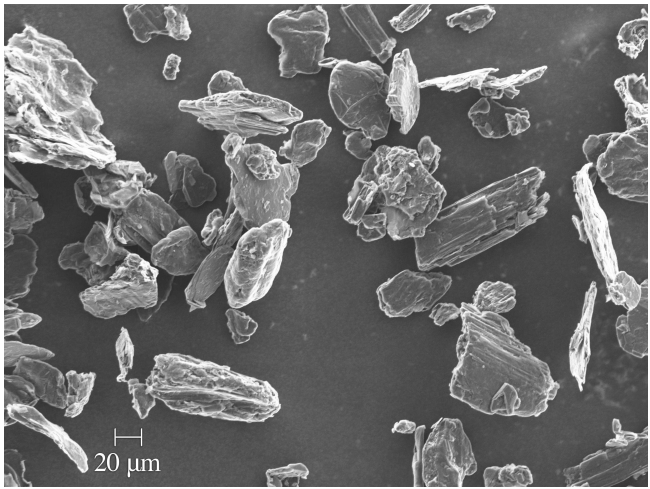}
    \caption{Scanning electron microscopy image of the Timcal Timrex KS5-75TT graphite particles used in this study obtained using a Zeiss EVO LS15 with an accelerating voltage of $\SI{10}{\kilo\volt}$.}
    \label{fig:methods:SEM:flake-graphite}
\end{figure}

Timcal Timrex KS5-75TT graphite particles with a size distribution of $D_{10}= \SI{9.1}{\micro\meter}$, $D_{50}=\SI{38.8}{\micro\meter}$, and $D_{90}=\SI{70.0}{\micro\meter}$ (supplier's laser diffraction data) were obtained from Imerys Graphite \& Carbon. They have a Scott (bulk) density of 0.44 $\si{g \ cm^{-3}}$ and a tapped density of 2.2 $\si{g \ cm^{-3}}$, see \ \cite{AMIDON2017271} for definitions, and are predominately flake-like with an irregular surface. However, included are also more oblate- and needle-shaped particles, Figure \ref{fig:methods:SEM:flake-graphite}. Suspensions at different volume fractions, $(\phi =$ 0.10, 0.15, 0.18, 0.20, 0.23, 0.25, 0.27, 0.28, 0.29), were prepared by mechanically mixing heavy mineral oil (Merck 330760; viscosity $\eta_0 = \SI{0.14}{\pascal\second}$ at \SI{25}{\celsius}) with graphite powder. In between experiments the suspensions were stored at room temperature.

For rheological measurements an AR-G2 rheometer (TA Instruments) was used, and temperature (\SI{10}{\celsius} to \SI{35}{\celsius}) was controlled with a Peltier system. A parallel plate geometry (diameter \SI{40}{\milli\meter}, gap height 600-\SI{1000}{\micro\meter}) with cross-hatching was used to obtain data for suspensions, while for the pure oil a smooth parallel plate geometry (same diameter, gap height \SI{500}{\micro\meter}) was used. In a parallel plate geometry of radius $R$ the imposed angular velocity, $\Omega$, leads to the reported rim shear rate $\dot{\gamma} = \Omega R/H$, with $H$ being the gap height. The measured torque, $M$, gives the corrected stress $\sigma = \left(M/2\pi R^3 \right) \cdot \left(3 + \mathrm{d}\ln{M}/\mathrm{d}\ln{\Omega} \right)$, and the relative viscosity is given as $\eta_r = \eta/\eta_0 = \sigma/\dot{\gamma}\eta_0$ \cite{Macosko1994}. Suspensions were mechanically homogenised prior to loading. Samples at $\phi < 0.27$ were pre-sheared for \SI{30}{\second} at a shear rate $\dot{\gamma} = \SI{10}{\second^{-1}}$ to erase memory of the loading history. At $\phi \geq 0.27$ we presheared for \SI{30}{\second} at $\sigma = \SI{150}{\pascal}$ or $\dot{\gamma} = \SI{0.5}{\second^{-1}}$ since $\dot{\gamma} = \SI{10}{\second^{-1}}$ caused visible sample fracture (see Figure S1), which for our samples occurs at $\sigma_{\rm max} = \SI{160}{\pascal}$. The duration of 30 s was found to be adequate to reach a stable viscosity, Figure S2. Before starting the experiments, samples equilibrated to the measurement temperature for \SI{3}{\minute} during which a steady state was reached as found by small amplitude oscillatory shear time sweeps, Figure S3. Steady shear measurements were performed on the $\phi = 0.10, \ 0.15, \ 0.18, \ 0.20, \ 0.23, \ 0.25$ suspensions to study time-temperature superposition, with gap heights of $600$, $800$, and \SI{1000}{\micro\meter} at each  temperature to check that wall slip effects were insignificant \cite{Wei2015,Blanc2018,Khalkhal2011}. We sheared from low to high rates with a time interval of 30 s between points. We checked that steady-state stresses from step-rate transients yielded similar flow curves, and the flow curves were independent of sweep direction for non-fractured samples (see Figures S4 and S5, respectively). Reported flow curves are the average of at least two measurements.

Brownian motion and inertia effects can be neglected because $\mbox{Pe} \gg 1$ and $\mbox{Re} < 10^{-3}$ throughout our range of $\dot\gamma$. The minimum stress required for counteracting sedimentation is estimated as $\sigma_{\rm min} = \Delta\rho g D$ \cite{Richards2020,Richards2021}, where $\Delta\rho$ is the particle fluid density difference, $g$ is the acceleration of gravity, giving $\sigma_{\rm min} \approx \SI{0.5}{\pascal}$ using $D=D_{50}$. This is similar to the lowest measured stresses in the two least concentrated suspensions, so that we may also neglect sedimentation effects.

\section{Results and Discussion}

\subsection{Temperature dependence}

\begin{figure}
    \includegraphics{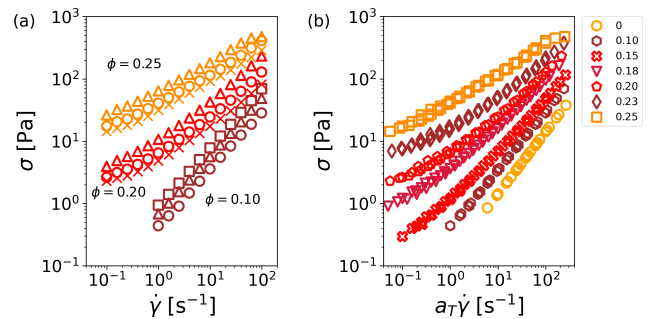}
    \caption{(a) Stress $\left(\sigma\right)$ as a function of shear rate $\left(\dot{\gamma} \right)$ at $T = \SI{10}{\celsius}$ $\left(\square \right)$, 
\SI{15}{\celsius} $\left(\triangle \right)$, \SI{25}{\celsius} $\left(\circ \right)$ and \SI{35}{\celsius} $\left(\times \right)$. (b) Time-temperature superposition of stress as a function of shifted shear rate $\left(a_T \dot{\gamma} \right)$ at a reference temperature of $25 \ ^{\circ}$C.}
    \label{fig:results:TTS_shifted_non-shifted}
\end{figure}

\begin{figure}
    \includegraphics{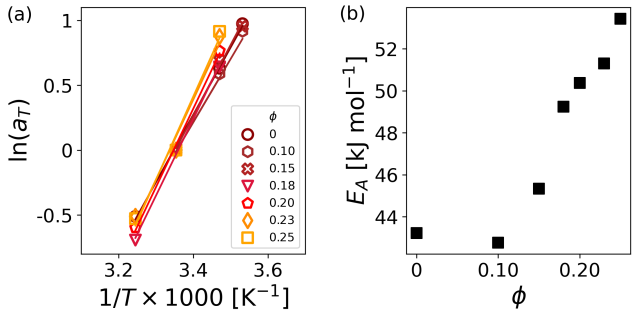}
    \caption{(a) Shift factors $\left(a_T \right)$, used for time-temperature superposition of the data in Figure \ref{fig:results:TTS_shifted_non-shifted}b, as a function of inverse temperature $\left(1/T\right)$. The natural logarithm of the shift factors follow an Arrhenius-dependence (solid lines), Equation \ref{eq:results:Arrhenius_Ea}. (b) Flow activation energies $E_A$ as a function of graphite volume fraction $\phi$, calculated from the Arrhenius-model fits in (a). $E_A$ exhibits a dependence on the graphite volume fraction.}
    \label{fig:results:stress_shift-factors}
\end{figure}

The flow curves of selected suspensions with $\phi \leq 0.25$ at \SI{10}{\degreeCelsius} to \SI{35}{\degreeCelsius} are shown in Figure \ref{fig:results:TTS_shifted_non-shifted}a. A common approach to analysing temperature-dependent rheology is time-temperature superposition (TTS), where (typically) the shear rates are shifted by some temperature-dependent factors $a_T$ to align the flow curves to that measured at some reference temperature. We shift our flow curves $\sigma(\dot\gamma,T)$ by the scaling $\dot\gamma \to a_T\dot\gamma$, Figure \ref{fig:results:TTS_shifted_non-shifted}b, to arrive at a set of concentration-dependent shift factors $a_T(\phi)$, Figure \ref{fig:results:stress_shift-factors}a. These follow an Arrhenius-type dependence,
\begin{equation}
    \ln{a_T(\phi)} = \frac{E_A(\phi)}{R} \left(\frac{1}{T} - \frac{1}{T_{\rm ref}} \right),
    \label{eq:results:Arrhenius_Ea}
\end{equation}
where $E_A$ is the flow activation energy, $R$ the universal gas constant, and  $T_{\rm ref}=\SI{25}{\degreeCelsius}$ the reference temperature. Plotting $E_A(\phi)$, Figure \ref{fig:results:stress_shift-factors}b, reveals a rapid increase at $0.15 \lesssim \phi \lesssim 0.20$. Interestingly, $\phi = 0.20$ is the first concentration at which we may discern a flattening at low $\sigma$ in the flow curve as $\phi$ increases, Figure \ref{fig:results:TTS_shifted_non-shifted}b, and therefore the emergence of a finite yield stress. 

TTS has been applied to a variety of non-Brownian suspensions, including silica in ethylene glycol/glycerol with added salt  \cite{Watanabe1996}, glass spheres in polyisobutylenes \cite{Schmidt2001} and PMMA spheres (diameter $>\SI{1}{\micro\meter}$) in poly($\epsilon$-caprolactone) \cite{Constanzo2019}. In these cases with Newtonian solvents, the suspended solids do not significantly affect the relaxation dynamics and hence shift factors. In cases where the matrix is non-Newtonian, there is more variation from system to system, with little to no change with filler concentration in glass fiber reinforced polypropylene \cite{Mobuchon2005} or Newtonian epoxy resin filled with multi-walled carbon nanotubes (MWCNTs) \cite{Khalkhal2011}, while the activation energy was found to increase with concentration in nitrile-butadiene/graphene nanocomposites \cite{Mowes2014} and decrease with concentration in MWCNT/polycarbonate composites \cite{Abbasi2009}. 

We will later apply the adhesive constraints model reviewed earlier to our system, which assumes a purely stress-dependent rheology, i.e., $\eta(\sigma) = \eta_0 \mathcal{F}(\phi,\sigma)$ for some function $\mathcal{F}(\phi,\sigma)$. In this case, we have
\begin{equation}
\frac{\sigma}{\eta_0(T)\dot\gamma}=\mathcal{F}(\phi,\sigma) = \frac{\sigma}{\eta_0(T_{\rm ref})a_T \dot\gamma} \Rightarrow a_T = \eta_0(T)/\eta_0(T_{\rm ref}),
\label{eq:results:TTS_susp}
\end{equation}
which we recognise as the shift factor for the pure solvent. So, we predict a $\phi$-independent suspension shift factor, as was observed in many previous systems~\cite{Watanabe1996,Schmidt2001,Constanzo2019}. Figure \ref{fig:results:stress_shift-factors}b therefore suggests an additional source of temperature-dependence beyond the solvent in our graphite-in-oil suspensions. 

\subsection{Phenomenological fitting}

\begin{figure}
  \includegraphics{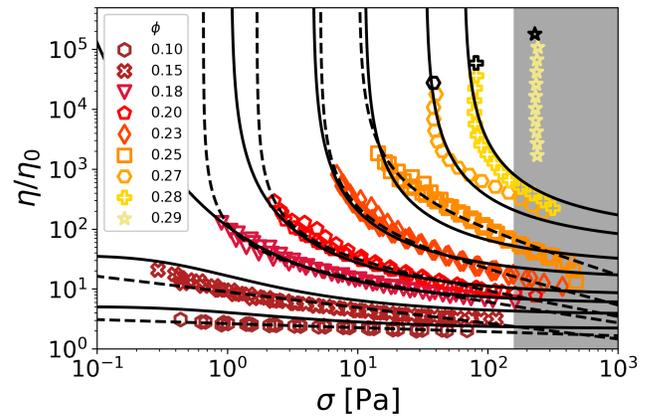}
  \caption{The relative viscosity $\left(\eta/\eta_0 \right)$ at $\SI{25}{\celsius}$ as a function of shear stress $\left(\sigma \right)$ at various $\phi$. The dashed lines show Herschel-Bulkley model fits, Equation \ref{eq:introduction:Herschel-Bulkley}, with parameters shown in Figure \ref{fig:results:HB_parameters}. The solid lines are the fits to the constraint-based model, Equation \ref{eq:results:Krieger-Dougherty_model}, Equation \ref{eq:introduction:ADH-phiJ} and Equation \ref{eq:introduction:a_sig}, with $l=2$, $\sigma_a = \SI{4.4}{\pascal}$, $\beta = 0.45$, $\phi_{\rm rlp} = 0.31$ and $\phi_{\rm alp} = 0.18$. Shaded area marks the region of sample fracture where the data are excluded from our fitting.}
  \label{fig:results:TTS:Constraint-rheology:viscosity-stress}
\end{figure}

Figure \ref{fig:results:TTS_shifted_non-shifted}b suggests that a finite yield stress emerges at $\phi \gtrsim 0.20$. Flow curves for yield-stress fluids are often fitted empirically by the Herschel-Bulkley (HB) equation  \cite{Herschel1926,Kalyon2014}
\begin{equation}
    \sigma = \sigma_y + K\dot{\gamma}^n,
    \label{eq:introduction:Herschel-Bulkley}
\end{equation}
where $\sigma_y$ is the HB yield stress, $K$ the consistency index, and $n$ the flow index ($n>1$ for shear thickening, $n<1$ for shear thinning).  This equation, which fits data from disparate systems from colloidal gels and glasses through jammed emulsions \cite{Bonn2017} to certain non-Brownian suspensions \cite{Mueller2010,Papadopoulou2020}, also credibly applies to our data for $\phi \leq 0.25$, Figure \ref{fig:results:TTS:Constraint-rheology:viscosity-stress}. 

\begin{figure*}
    \includegraphics{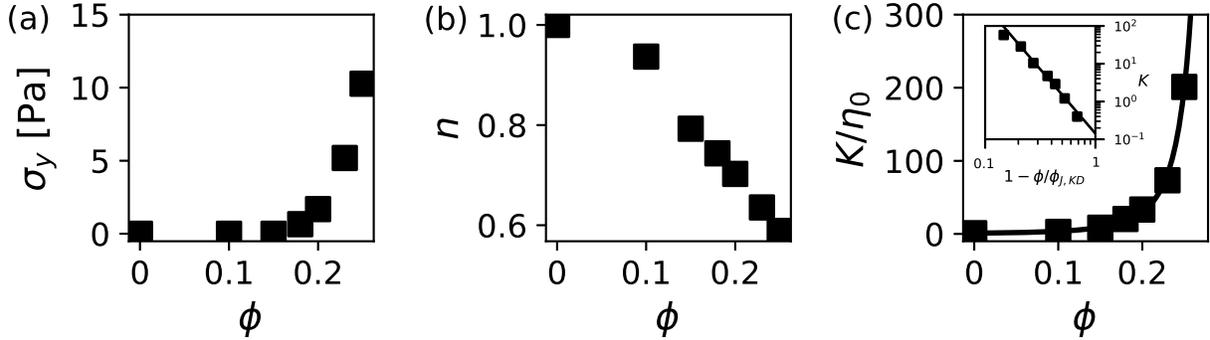}
    \caption{Parameters from the Herschel-Bulkley fit to the time-temperature superposition data, Figure \ref{fig:results:TTS:Constraint-rheology:viscosity-stress}, as a function of the graphite volume fraction $\phi$. (a) Yield stress, $\sigma_y$. (b) Flow index, $n$. (c) Normalized consistency index, $K/\eta_0$, with the Krieger-Dougherty model (solid line) from Equation \ref{eq:results:Krieger-Dougherty_model} fitted to the data using $l_{KD} = 3.40$ and $\phi_{J,KD} = 0.32$. The inset shows the same data as the main figure but with $K$ plotted as a function of $1 - \phi/\phi_{J,KD}$.}
    \label{fig:results:HB_parameters}
\end{figure*}

The fitted HB yield stress $\sigma_y$ is plotted as a function of $\phi$ in Figure \ref{fig:results:HB_parameters}a, where we have assigned $\sigma_y = 0$ to all systems with $\phi \leq 0.15$. This reflects the lack of a low-$\sigma$ plateau in the relevant flow curves, Figure \ref{fig:results:TTS_shifted_non-shifted}, and is consistent with our experience of handling these samples, which flowed easily as liquids. The extracted HB yield stress increases rapidly for $\phi \gtrsim 0.20$, Figure \ref{fig:results:HB_parameters}a, suggesting a divergence at some critical concentration.

The flow index $n$ steadily decreases with increasing graphite volume fraction, Figure \ref{fig:results:HB_parameters}b,  from $n = 1$  for our Newtonian mineral oil ($\phi=0$) to $n\approx 0.6$ for $\phi=0.25$, indicating an increasing degree of shear thinning. This contrasts with yield stress suspensions of spherical particles, where the $\phi$ dependence is either weaker \cite{Mueller2010} or absent \cite{Papadopoulou2020}. Our observation possibly reflects a particle shape effect. 

Like $\sigma_y$, the consistency index $K$ also increases with $\phi$ and appears to diverge, Figure \ref{fig:results:HB_parameters}c. The data can be fitted to $K\propto(1-\phi/\phi_{J,KD})^{-l_{KD}}$ with $\phi_{J,KD} = 0.32$ and $l_{KD} = 3.40$ (KD = Krieger-Dougherty). The consistency index scales as the viscosity at $\dot\gamma = \SI{1}{\second^{-1}}$ and it may be interpreted as a kind of `generalised viscosity'. Perhaps unsurprisingly, $\phi_{J,KD}$ will be related to a viscosity divergence to be identified later. 
\subsection{Physical modelling}

The HB equation is purely phenomenological with the parameters $K$ and $n$ containing no clear physical meaning. In contrast, a recently-proposed constraint-based model~\cite{Guy2018} is predicated on the existence of adhesive and frictional contacts. Fitting to this model may therefore yield insights into the micro-physics of our graphite suspensions. 

Since our flow curves all show shear thinning $\left(d\eta_r/d\sigma < 0 \right)$ without subsequent shear thickening, Figure \ref{fig:results:TTS:Constraint-rheology:viscosity-stress}, we apply the model in the limit where  particle contacts are frictional at all stresses ($f=1$)\footnote{The absence of any shear thickening implies the frictional contact state is unchanged within our measured stress range (up to \SI{500}{\pascal}), indicating $f$ should be fixed at either 0 (always friction-less) or 1 (always frictional). Given that our flake graphite has not been stabilised, we should expect that $\sigma^*\to 0$ and follow the convention in refs. \ \cite{Richards2020,Richards2021} describing similar un-stabilised systems and set $f=1$. We note that the model can technically also describe the $f=0$ case with the replacements $\phi_{\rm rlp} \to \phi_{\rm rcp}$ and $\phi_{\rm alp} \to \phi_{\rm acp}\simeq\phi_{\rm rlp}$ in Eq. \ref{eq:introduction:ADH-phiJ}, see refs. \ \cite{Guy2018,Richards2020})}. Adhesive  constraints to rolling that are progressively released as stress increases give rise to yielding and subsequent shear thinning. This is similar to suspensions of cornstarch in oil \cite{Richards2020} and unstabilised calcite in glycerol-water  \cite{Richards2021}. 

The model, which, has been reviewed above, requires five parameters: $l$ the viscosity divergence exponent, two critical volume fractions $\phi_{\rm rlp}$ and $\phi_{\rm alp}$ reflecting the frictional and frictional+adhesive jamming points, and the stress scale $\sigma_a$ and exponent $\beta$ describing the release of adhesive constraints. If particle contacts do not change with $\phi$, a single set of $\phi$-independent parameters should fit all our flow curves. This is in contrast to the HB fits, with either two or three free parameters per flow curve. 

Sample fracture was observed at $\sigma > \SI{160}{\pascal}$, so that the data in the shaded region in Figure \ref{fig:results:TTS:Constraint-rheology:viscosity-stress} are  not used for fitting. Fitting to the rest of the data gives $\sigma_a = \SI{4.4}{\pascal}$ for the release of adhesive rolling constraints, with an exponent $\beta = 0.45$. The latter is close to previous work that found $0.5 \lesssim \beta \lesssim 0.6$ \cite{Guy2018,Richards2021}, evidencing a similar rate of release of rolling constraints.

Our fitted value of $\phi_{\rm rlp} = 0.31$ is close to $\phi_{J,KD} = 0.32$ obtained from the consistency index data from fitting to the KD equation, Figure \ref{fig:results:HB_parameters}c. This concentration is where friction alone suffices to cause jamming, and is well below the random loose packing limit for monodisperse frictional spheres, even in the limit of infinite static friction, where $\phi_{\rm rlp} \approx 0.55$ \cite{Silbert2010}. Interestingly, this latter value appears not particularly dramatically sensitive to details of particle morphology, as fitting the same model to suspensions of more or less isotropic calcite crystals also returns $\phi_{\rm rlp} \approx 0.55$ \cite{Richards2021}. However, gross changes in morphology do have an effect. Fitting $K(\phi)$ in suspensions of prolate wollastonite with a mean aspect ratio of 9 suspended in silicone oil returned $\phi_{\rm rlp} = 0.32$-0.34 \cite{Mueller2010}, which is close to the value for our suspensions of oblate and needle-shaped graphite particles. While polydispersity will increase $\phi_{\rm rlp}$  \cite{doi1988,Farr2009}, in our case, the non-spherical particle morphology apparently completely counteracts this effect.

\begin{figure}
    \includegraphics{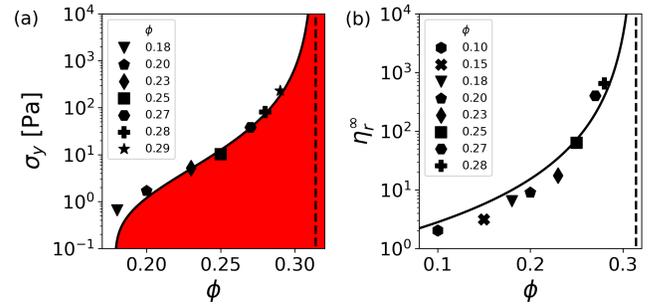}
    \caption{(a) Yield stresses $\left( \sigma_y \right)$ from Figure \ref{fig:results:HB_parameters}a, including $\sigma_y\left(\phi = 0.27 \right) = 38.3$ Pa, $\sigma_y\left(\phi = 0.28 \right) = 81.3$ Pa, and $\sigma_y\left(\phi = 0.29 \right) = 230$ Pa (black symbols in Figure \ref{fig:results:TTS:Constraint-rheology:viscosity-stress}) measured at the lowest shear rate. Data are consistent with predictions by Equation \ref{eq:introduction:sigY} (solid line) using the parameters found by fitting the flow curves in Figure \ref{fig:results:TTS:Constraint-rheology:viscosity-stress}. Standard deviations were comparable to or smaller than the size of the symbols and were hence left out. Shaded (red), jammed; unshaded, flowing. Dotted line, $\phi_{\rm rlp} = 0.31$. (b) Relative viscosity values before fracture $\left(\eta_r^{\infty} \right)$ fitted to $\eta_r^{\infty} = \left(1 - \phi/\phi_{\rm rlp} \right)^{-l}$ using $l = 2.69$ and $\phi_{\rm rlp} = 0.31$ (dotted vertical line).}
    \label{fig:results:TTS:Constraint-rheology}
\end{figure}

Equation \ref{eq:introduction:sigY} is one of the key predictions of the adhesive constraint model. It shows that a finite yield stress should only emerge at $\phi_{\rm alp}$; thereafter, $\sigma_y$ increases until it diverges at $\phi_{\rm rlp}$. To test this prediction,  we collected flow curves at three additional higher volume fractions, $\phi = 0.27, 0.28$ and 0.29, Figure \ref{fig:results:TTS:Constraint-rheology:viscosity-stress}, to obtain $\sigma_y(\phi)$ over as wide a range of $\phi$ as possible. The yield stresses for these three volume fractions were estimated as the stress at the lowest accessed shear rate (sometimes called the `apparent yield stress'~\cite{Barnes1999}). These three yield stresses are plotted together with those obtained by fitting the HB equation in Figure \ref{fig:results:TTS:Constraint-rheology}a. Equation \ref{eq:introduction:sigY}  (the full line) with the constraint-model parameters from Figure \ref{fig:results:TTS:Constraint-rheology:viscosity-stress} is seen to be consistent with this data. Below $\phi_{\rm alp} \approx 0.18$, adhesive contacts are insufficient to stabilise frictional force chains~\cite{Richards2020} so that a non-flowing state never forms -- there is no yield stress. At $\phi_{\rm rcp} \approx 0.31$, frictional force chains are stable irrespective of adhesion, and the system is permanently jammed. The latter is also evidenced by a plot of the high-shear viscosity against volume fraction, Figure \ref{fig:results:TTS:Constraint-rheology}b, which shows a divergence at a concentration that is consistent with $\phi_{\rm rlp}$. 

It is intriguing that our fitted value of $\phi_{\rm alp} \approx 0.18$ is comparable to that found for suspended particles of adhesive calcite \cite{Richards2021} and PMMA \cite{Guy2018} ($\approx 0.18$ and 0.20 respectively) even though our fitted value of $\phi_{\rm rlp} \approx 0.31$ is considerably below the value of $\approx 0.55$ obtained for these two systems. Intuitively, one may expect adhesive loose packing to correlate with contact percolation. Particle clusters in sheared non-colloidal suspensions may form a percolated network around a volume fraction of $0.3 - 0.4$ for monodisperse spheres \cite{Gallier2015}, with lower thresholds for anisotropic particles. So, carbon nanotubes may percolate at as low as 0.3-0.8~wt.\% in polycarbonate \cite{Abbasi2009}, corresponding to $\phi \lesssim 0.005$. Similarly, the electrical percolation threshold of polymer composites decreases with filler aspect ratio \cite{Antunes2011}. We may therefore expect $\phi_{\rm alp}$ for our flake-like graphite particles to be lower, if not significantly lower, than that in adhesive PMMA spheres~\cite{Guy2018} or calcite crystals that do not show gross geometric anisotropy~\cite{Richards2021}. 
These considerations suggest that contact percolation alone is insufficient for adhesive loose packing. This is expected from the physical picture associated with the adhesive constraint model~\cite{Richards2021}, that $\phi_{\rm alp}$ is the lowest concentration at which adhesive contacts can stabilise frictional force chains to cause jamming. In other words, these contacts do {\it not} act alone, so that percolation {\it per se} is insufficient, explaining the observation that the fitted $\phi_{\rm alp} \approx 0.18$ in our graphite suspensions is considerably higher than what one may expect to be needed for contact percolation alone. 

\begin{figure*}
    \includegraphics{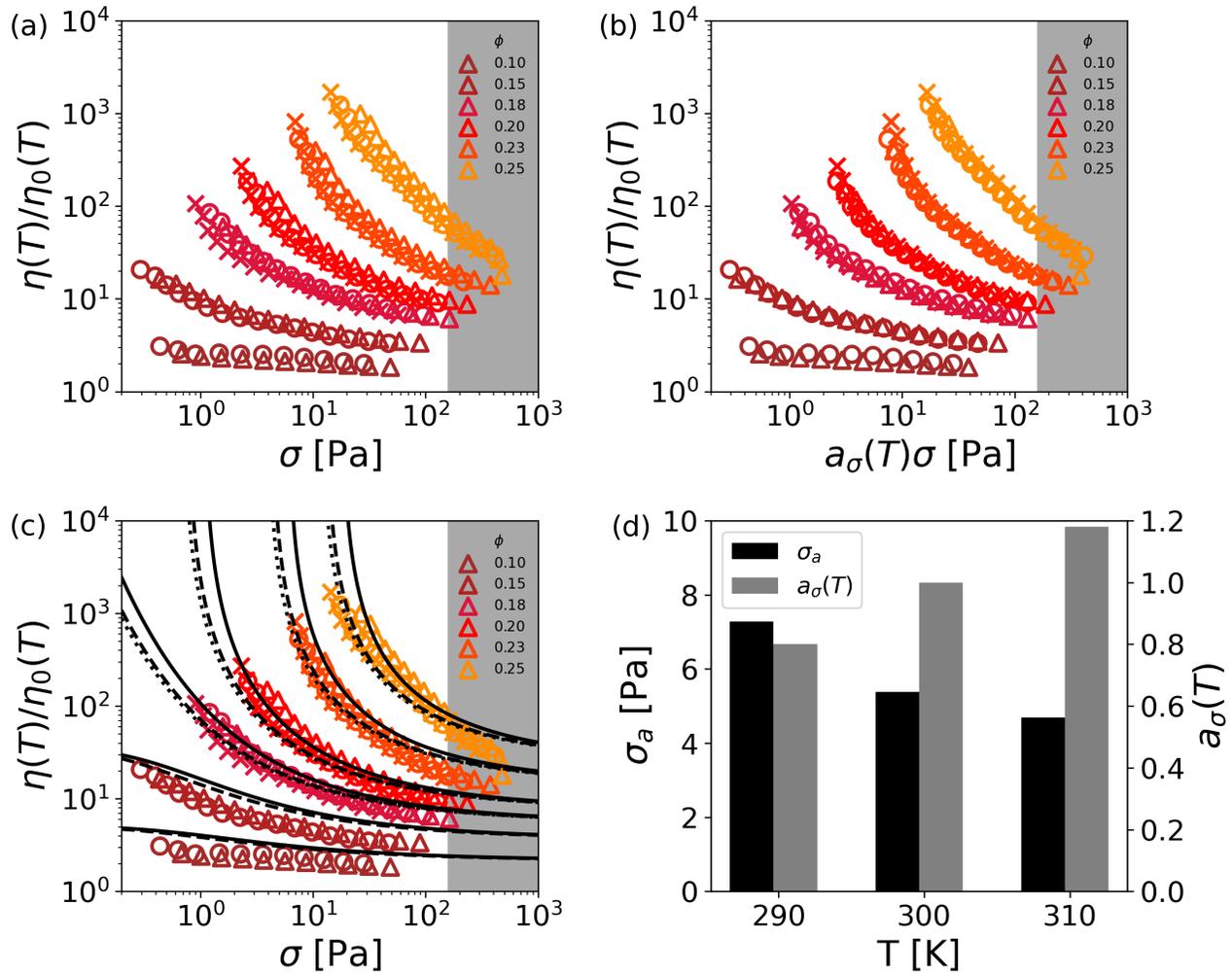}
    \caption{(a) Relative viscosities of suspensions with various filler volume fractions $\phi$ at temperatures of $15 \ ^{\circ}$C $\left(\triangle \right)$, $25 \ ^{\circ}$C $\left(\circ \right)$ and $35 \ ^{\circ}$C $\left(\times \right)$. (b) Same as (a) but stress values for all suspensions at a given temperature have been shifted by a factor $a_{\sigma}(T)$ (see (d)) to collapse all data onto the curves obtained at $25 \ ^{\circ}$C. (c) Same as (a) but with the constraint-model, Equation \ref{eq:introduction:WC-phiJ} and  Equation \ref{eq:introduction:ADH-phiJ}, fitted to data at each temperature (full line $15 \ ^{\circ}$C; dashed line $25 \ ^{\circ}$C; dotted line $35 \ ^{\circ}$C) using $\beta = 0.38$, $\phi_{\rm rlp} = 0.31$, $\phi_{\rm alp} = 0.18$. (d) Values of $\sigma_a(T)$ used in the constraint-model fits in (c). Shaded (grey) area in (a)-(c) marks the region of stresses in which sample fracture was observed.}
    \label{fig:results:Relative_viscosity_stress}
\end{figure*}

Finally, we return to the unexpected temperature-dependent flow behaviour, which manifested as a sharp increase in the TTS shift factors above $\phi =0.15$, Figure \ref{fig:results:stress_shift-factors}b.  Plotting the relative viscosities $\eta/\eta_0(T)$ at $15 \ ^{\circ}$C, $25 \ ^{\circ}$C and $35 \ ^{\circ}$C against the shear stress, we see that data collapse becomes progressively worse as $\phi$ increases, and becomes very noticeable for $\phi > 0.15$, Figure \ref{fig:results:Relative_viscosity_stress}a. However, we find that shifting the {\it stress} by a temperature-dependent factor, $a_{\sigma}(T)$, improves the data collapse noticeably, especially at $\phi \geq 0.15$, Figure \ref{fig:results:Relative_viscosity_stress}b. Fitting the resulting flow curves to the adhesive constraint model, Figure \ref{fig:results:Relative_viscosity_stress}c, using 
a fixed $\beta$-value, $\phi_{\rm rlp} = 0.31$ and $\phi_{\rm alp} = 0.18$ then suggests a temperature-dependent characteristic adhesive stress $\sigma_a=\sigma_a(T)$, which, as expected, is anti-correlated with the temperature-dependent shift factor $a_\sigma(T)$, Figure \ref{fig:results:Relative_viscosity_stress}d.

The finding that $\sigma_a$ decreases with temperature suggests rolling constraints are more easily released at higher temperatures. The origin of this effect in our system is currently unobvious. The van der Waals interaction should be temperature-independent in our temperature range~\cite{Israelachvili1992}, so that there is another source of rolling constraint that is more specific to our graphite suspensions. It would be fruitful to examine whether solvent-mediated interactions, such as improved wetting driven by a decreased oil surface tension with temperature could be the source of this temperature dependence. Whatever its source, a temperature-dependent $\sigma_a$ provides a handle for controlling the rheology of our system. (Note that such temperature variation in $\sigma_a$ does not change the critical packing limits in Equation \ref{eq:introduction:sigY}, only the relative magnitude of the yield stress.)

\section{Conclusion}

Polydisperse flake-like graphite particles suspended in heavy mineral oil were investigated using rotational rheometry. Time-temperature superposition of the flow curves revealed an Arrhenius-type dependence of the horizontal shift factors with an increasing flow activation energy with filler concentration above a threshold graphite volume fraction $\phi \approx 0.15$. 

We were able to account for our observations by fitting to a recent constraint-based model of non-Brownian suspension rheology~\cite{Richards2020} in which frictional and adhesive contacts constrain sliding and rolling motion, respectively. Above a critical concentration $\phi_{\rm alp}$, adhesive stabilisation of frictional force chains gives rise to a finite yield stress, which diverges at a higher $\phi_{\rm rlp}$, at which frictional contacts alone suffice for mechanical stability. At all $\phi < \phi_{\rm rlp}$, stress-induced release of adhesive rolling constraints causes shear thinning, allowing particles to align themselves in the flow direction.

Our analysis suggests a temperature-dependent characteristic adhesive stress $\sigma_a=\sigma_a(T)$ that decreases with temperature. Further investigations are required to determine the underlying cause of this observation. The adhesive stress is suggested to be related to a critical torque $\sigma_a \propto M^*/R^3$ to peel apart adhesive contacts between particles with a mean size $R$~\cite{Guy2018}, so that the temperature-dependence we observe should also manifest in single-particle measurement of the contact tribology \cite{Heim1999}.

Finally, and more generally, that the adhesive constraint model~\cite{Richards2020} can give a credible account of data for PMMA spheres in oil \cite{Guy2018}, cornstarch in oil \cite{Richards2020}, calcite in glycerol-water \cite{Richards2021} and molten chocolate (= sugar grains in oil)~\cite{Blanc2018}, and now, plate-like graphite in oil, suggests that the micro-physical basis of this model may be sound. In particular, our finding that $\phi_{\rm alp} \approx 0.18$ is significantly higher than what one might expect from contact percolation alone in flake-like particles provides support for the suggestion~\cite{Richards2020,Richards2021} that yielding behaviour is due to the combined action of adhesive contacts stabilising frictional force chains. Our work therefore illustrates the advantage of fitting data using such a model compared to a purely phenomenological equation such as the Herschel-Bulkley model. A physics-based model permits inference about microscale mechanisms, such as temperature-dependent adhesive contacts, which are susceptible to further experimental testing.

\section{Supplementary Material}

The supporting information contains additional details related to the experimental methods applied in this work, including evidence of sample fracture, flow curve dependence on sweep direction, and sample equilibration after shear cessation.  

\begin{acknowledgments}
This work was financially supported by Advent Technologies A/S and a grant from the Industrial PhD programme, Innovation Fund Denmark, project 8053-00063B.
\end{acknowledgments}

\section*{Data Availability Statement}

The data that support the findings of this study are available from the corresponding author upon reasonable request.


%
%

%


\bibliography{bibliography}

\end{document}